\documentclass[sigconf,nonacm,screen=true,bookmarks=false,9pt]{acmart}
\iftrue
\fi

\usepackage{tablefootnote} 
\usepackage{graphicx}

\usepackage{amsmath}
\usepackage{mathtools}
\usepackage{comment}
\usepackage[subrefformat=parens,labelformat=parens]{subfig}
\usepackage{bm}
\usepackage{multirow}
\usepackage{diagbox}
\usepackage{multicol}
\usepackage{booktabs}
\usepackage{placeins}
\usepackage{graphicx}
\usepackage{tikz}
\usepackage{balance}
\usepackage{courier}                                       
\usepackage{cleveref}                                      
\usepackage[mathcal]{eucal}
\usepackage[noend]{algpseudocode}
\algrenewcommand\textproc{\texttt}
\makeatletter\let\float@addtolists\relax\makeatother
\usepackage{algorithm}
\usepackage{filecontents}                                  
\usepackage{pgfplots}
\usepackage{pgfplotstable}
\pgfplotsset{compat=newest}
\usepackage{siunitx}
\theoremstyle{plain}

\theoremstyle{definition}

\usepackage[skip=2pt]{caption}
\renewcommand{\abstractname}{\MakeUppercase{Abstract}}
\renewcommand{\refname}{\MakeUppercase{References}}

\graphicspath{{./figs/}}
\usepackage{siunitx}
\setcopyright{none}
\acmDOI{}
\acmISBN{}

\begin{document}

\title{SPLIT-MOR: Fast and Accurate Model Order Reduction for Nonlinear Transient Thermal Simulation of 2.5D/3D ICs}

\author{Zebin Shi}
\affiliation{%
  \department{School of Physics}
  \institution{Peking University}
  \city{Beijing}
  \country{China}}
\email{shizebin@stu.pku.edu.cn}

\author{Tianxiang Zhu}
\affiliation{%
  \department{School of Integrated Circuits}
  \institution{Peking University}
  \city{Beijing}
  \country{China}}
\email{txzhu@pku.edu.cn}

\author{Yibo Lin}
\authornote{Corresponding authors.}
\affiliation{%
  \department{School of Integrated Circuits}
  \institution{Peking University}
  \city{Beijing}
  \country{China}}
\affiliation{%
  \department{Institute of Electronic Design Automation}
  \institution{Peking University}
  \city{Wuxi}
  \country{China}}
\email{yibolin@pku.edu.cn}

\author{Runsheng Wang}
\authornotemark[1]
\affiliation{%
  \department{School of Integrated Circuits}
  \institution{Peking University}
  \city{Beijing}
  \country{China}}
\affiliation{%
  \department{Institute of Electronic Design Automation}
  \institution{Peking University}
  \city{Wuxi}
  \country{China}}
\email{r.wang@pku.edu.cn}

\renewcommand{\shortauthors}{Shi et al.}

\makeatletter
\def\@mkauthors@iii{%
  \gdef\@currentauthors{}%
  \global\setbox\mktitle@bx=\vbox{%
    \box\mktitle@bx\par\medskip
    \centering
    {\Large
      Zebin Shi\textsuperscript{1},
      Tianxiang Zhu\textsuperscript{2},
      Yibo Lin\textsuperscript{2,3,*},
      Runsheng Wang\textsuperscript{2,3,*}\par}
    \smallskip
    {\large
      \{\textsuperscript{1}School of Physics, \textsuperscript{2}School of Integrated Circuits\}, Peking University, Beijing, China\par
      \textsuperscript{3}Institute of Electronic Design Automation, Peking University, Wuxi, China\par}
    {\normalsize
      Email: \nolinkurl{shizebin@stu.pku.edu.cn},
      \{\nolinkurl{txzhu}, \nolinkurl{yibolin}, \nolinkurl{r.wang}\}\nolinkurl{@pku.edu.cn}\par}
    \bigskip}}
\makeatother

\begin{abstract}
Fast and accurate transient thermal simulation is essential to thermal reliability analysis and dynamic thermal management of advanced packages, and model order reduction (MOR) techniques are widely applied to overcome the efficiency bottleneck of rigorous numerical methods. However, nonlinear terms including temperature-dependent leakage power, thermal conductivity, and heat capacity prevent the application of conventional linear MOR, while hyper-reduction solutions require costly nonlinear training data. In this work, we propose \textit{SPLIT-MOR} (\underline{S}ubdomain-wise \underline{P}iecewise \underline{L}inearization with \underline{I}nterval \underline{T}ensors), a novel nonlinear MOR method that requires no additional training set. \textit{SPLIT-MOR} computes reduced component tensors by subdomain clustering and temperature-interval partitioning in the offline stage, and updates the reduced coefficient matrices and load vectors efficiently in the online stage. Experimental results show that the online solving time of \textit{SPLIT-MOR} achieves up to an $830\times$ speedup over COMSOL and a $7.5\times$ speedup over the state-of-the-art hyper-reduction method MDEIM, with a $27\times$ total reduction in the offline time. The relative average error remains below $\qty{0.11}{\percent}$, and the maximum error is below $\qty{0.86}{\degreeCelsius}$, demonstrating the high accuracy of our method.
\end{abstract}

\maketitle
\pagestyle{plain}

\section{\MakeUppercase{Introduction}}
\label{sec:intro}

As integrated circuits approach their physical limits in the post-Moore era, advanced packaging technologies such as 2.5D/3D integration have become important paths to improving system computing capability~\cite{lau2022recent,biswas2024stco,chen2025breaking,lee2023extremely,das2024bandwidth}. Their extremely high spatial integration sharply increases on-chip power density and thermal coupling. Thermal failure has therefore become a major bottleneck in the system reliability of 2.5D/3D ICs~\cite{moroz20243dic,zhu2025high}. On the one hand, accurate thermal-reliability evaluation of typical workloads at design time is essential for efficiently optimizing cooling layouts and preventing physical damage ~\cite{pathania2018hotsniper,siddhu2022comet}. On the other hand, dynamic thermal management (DTM) is rapidly developing as an intelligent temperature-control technique at run time~\cite{guo2019leakage,rapp2021smartboost}. Both scenarios require fast and accurate transient thermal simulation of the system. Meanwhile, the temperature-dependent nonlinearity of the leakage power and the material properties become increasingly significant as the typical temperature of the system rises. Neglecting these dependencies will yield inaccurate temperature results and mislead subsequent optimizations. Efficient and accurate nonlinear transient thermal simulation is consequently urgently needed.

\begin{figure}[t]
  \centering
  \includegraphics[width=0.45\textwidth]{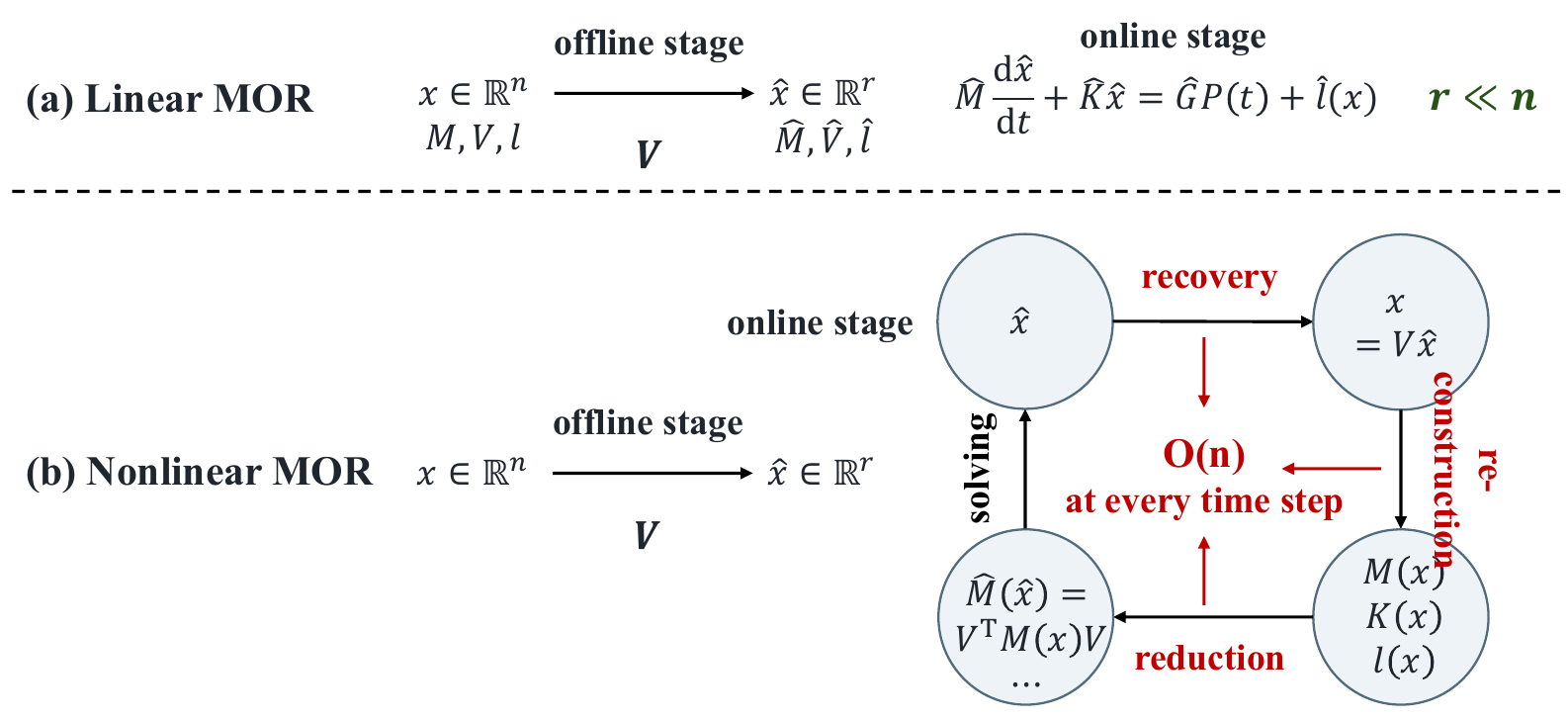}
  \caption{Computational flows of linear and nonlinear projection-based MOR. Linear MOR constructs fixed reduced coefficient matrices and load vectors offline, whereas nonlinear MOR repeatedly reconstructs the full-order matrices/vectors and projects them onto the reduced basis online, requiring full-order ($O(n)$) work at every time step.}
  \label{fig:intro}
\end{figure}

High-fidelity simulators based on finite-element methods (FEM) or finite-volume methods (FVM) have been widely used for transient thermal simulation of 2.5D/3D ICs~\cite{zhang2015hotspot,ladenheim2018mta,11539294}. Nevertheless, nonlinear transient simulation for large workloads can still take from hours to weeks. Model order reduction (MOR) methods significantly lower this cost by projecting the full-order system onto a low-dimensional space~\cite{xie2013system,benner2015survey,antoulas2005approximation,sirovich1987turbulence,odabasioglu1998prima} and have been widely used to accelerate transient thermal simulation, but two obstacles remain. First, MOR methods can be classified into data-driven or non-data-driven methods. Data-driven methods such as proper orthogonal decomposition (POD) require expensive full-order snapshots and may generalize poorly beyond their training regime, while classical non-data-driven methods such as moment matching and balanced truncation largely rely on linear time-invariant dynamics. Second, and more fundamentally, nonlinearity: with temperature-dependent leakage power, thermal conductivity, and heat capacity, the coefficient matrices and load vectors must be reassembled and re-projected at every Newton iteration, suffering from a full-order computational burden, as shown in Fig. \ref{fig:intro}.

Hyper-reduction techniques, including discrete empirical interpolation (DEIM), matrix DEIM (MDEIM), energy-conserving sampling and weighting (ECSW), and localized DEIM, are specially designed to alleviate this nonlinearity bottleneck~\cite{chaturantabut2010deim,wirtz2014deim,farhat2014dimensional,peherstorfer2014localized}. They recover the temperature field and update reduced coefficient matrices \& load vectors only at a small set of preselected sample points and their adjacent elements, thus eliminating the full-order complexity. Although they offer substantial acceleration potential, hyper-reduction methods are data-driven: expensive full-order snapshots are required to select interpolation points, which increases offline cost. Meanwhile, online reconstruction of nonlinear terms at the sample points still requires recovery over all degrees of freedom in adjacent elements. This cost is particularly evident in Jacobian construction, where the derivative of each degree-of-freedom temperature with respect to the reduced vector must be back-propagated, leaving room for further online time optimization. Neural network methods provide another kind of solution for nonlinear MOR~\cite{hesthaven2018nonintrusive,lee2020manifold,zhu2024fasttherm}. However, they heavily rely on abundant training data spanning the operating conditions, and their extrapolation behavior beyond the training regime is difficult to guarantee. These limitations of data-driven methods are particularly restrictive in current scenarios where nonlinear thermal snapshots are obtained through expensive full-order simulations and workload patterns vary. Non-data-driven structure-preserving projection has also been explored for nonlinear thermal simulation, but only for power-law temperature-dependent thermal conductivity~\cite{codecasa2014nonlinear}. This limits its applicability to more general material properties and heat-generation models.

To address these limitations, we propose \textit{SPLIT-MOR} (\underline{S}ubdomain-wise \underline{P}iecewise \underline{L}inearization with \underline{I}nterval \underline{T}ensors), a novel non-data-driven MOR method for fast nonlinear transient thermal simulation of 2.5D/3D ICs. The key idea is to replace per-iteration full-order work with a small, cached set of reduced component tensors selected online per subdomain. In the offline stage, \textit{SPLIT-MOR} pre-computes the component tensors by clustering thermally similar elements into subdomains and partitioning each nonlinear term into temperature intervals under an affine-error tolerance. In the online stage, it tracks each subdomain's active interval and updates the global coefficient matrices \& load vectors according to the selected component tensors only when a subdomain switches intervals, significantly increasing efficiency. The main contributions of our work are:
\begin{itemize}
  \item We propose a novel reduced component tensor technique based on subdomain clustering and temperature-interval partitioning, which removes the full-order computational burden from the online stage of nonlinear MOR without relying on an additional full-order snapshot training set, surpassing previous methods.
  \item We propose an efficient coefficient matrix \& load vectors update strategy that re-selects pre-computed reduced component tensors only for subdomains switching intervals in the online stage, further improving the online efficiency.
  \item We evaluate \textit{SPLIT-MOR} on two 2.5D and one 3D advanced packages. Its online solving time achieves up to an $830\times$ speedup over COMSOL and a $7.5\times$ speedup over the state-of-the-art hyper-reduction method MDEIM, and the total offline time is reduced by up to $27\times$, with relative average error below $0.11\%$ and maximum error below $\qty{0.86}{\degreeCelsius}$.
\end{itemize}

The rest of the paper is organized as follows. Section \ref{sec:preli} introduces necessary preliminaries. Section \ref{sec:splitmor} elaborates on the \textit{SPLIT-MOR} algorithm.
Section \ref{sec:experiments} presents the experimental results and discussions. Section \ref{sec:conclu}
concludes the paper. The source code has been released at~\cite{splitmor_code}.

\section{\MakeUppercase{Preliminaries}}
\label{sec:preli}

This section first formulates the full- and
reduced-order nonlinear thermal systems of 2.5D/3D ICs, then reviews how non-data-driven Krylov methods
construct reduced bases, and finally introduces the hyper-reduction techniques invented to deal with the nonlinear terms which cannot be efficiently handled by conventional linear MOR methods, together with their drawbacks.

\subsection{Thermal Model and Projection-Based MOR} \label{sec: mor}
We model an advanced package as a physical domain
$\Omega\subset\mathbb R^3$ with boundary $\partial\Omega$. At position
$\mathbf r\in\Omega$ and time $t$, let
$\theta(\mathbf r,t)=T(\mathbf r,t)-T_{\rm amb}$ denote the temperature
rise above the ambient temperature. Dynamic power density $p_d$ (external input) and
temperature-dependent leakage-power density $p_{\rm leak}$ (internally calculated) generate heat,
whereas the volumetric heat capacity $c_v$ and thermal conductivity
$k$ govern heat storage and diffusion. Allowing these quantities to
depend on temperature gives the nonlinear heat
equation:
\begin{equation}
c_v(\mathbf r,\theta)\partial_t\theta
-\nabla\!\cdot\![k(\mathbf r,\theta)\nabla\theta]
=p_d(\mathbf r,t)+p_{\rm leak}(\mathbf r,\theta)\,,
\label{eq:pde}
\end{equation}
subject to the initial condition
$\theta(\mathbf r,0)=\theta_0(\mathbf r)$ and the Robin boundary
condition $-k\nabla\theta\cdot\mathbf n=h\theta$. Here $\mathbf n$ is
the outward unit normal and $h$ is the heat-transfer coefficient;
$h=0$ recovers an adiabatic boundary.

The continuous model must be converted into an algebraic system for
numerical time integration. A finite-element discretization with $n$
temperature degrees of freedom (DoFs) gives:
\begin{equation}
M(x)\dot x+K(x)x=GP(t)+l(x)\,. \label{eq:fom}
\end{equation}
Here $x(t)\in\mathbb R^n$ is the nodal temperature-rise vector,
$M(x)$ and $K(x)$ are the temperature-dependent coefficient (heat-capacity and
conductivity) matrices, and $GP(t)$ and $l(x)$ are the load (dynamic power and leakage power) vectors. The input vector
$P(t)$ specifies the dynamic power at the thermal ports, and $G$ maps
these port inputs to the finite-element degrees of freedom.

Projection-based MOR reduces the dimension of the evolving state.
Let $V\in\mathbb R^{n\times r}$ be an orthonormal basis with $r\ll n$.
Approximating $x\approx V\widehat x$ and applying Galerkin projection
to Eq.~\eqref{eq:fom} gives a reduced spatial representation:
\begin{equation}
\widehat M(\widehat x)\dot{\widehat x}+\widehat K(\widehat x)\widehat x=\widehat G P(t)+\widehat l(\widehat x)\,, \label{eq:rom}
\end{equation}
where $\widehat x(t)\in\mathbb R^r$ is the reduced state. The reduced
coefficient matrices are $\widehat M=V^\top M(V\widehat x)V$ and
$\widehat K=V^\top K(V\widehat x)V$; similarly,
$\widehat G=V^\top G$ and $\widehat l=V^\top l(V\widehat x)$. This
projection reduces the DoFs from dimension $n$ to $r$, but it does not by itself reduce the
cost of construction $M$, $K$, and $l$: their nonlinear dependence on
$V\widehat x$ still requires online work on the full-order DoFs at every time step because $\widehat x$ continuously updates. Thus, nonlinear MOR has two distinct requirements: a projection basis $V$ that represents the thermal
state accurately and an efficient way to evaluate the projected
nonlinear operators. We discuss the first requirement next.

\subsection{Reduced-Basis Construction}
The moment matching method is a widely adopted linear MOR method in constructing the projection basis $V$~\cite{antoulas2005approximation,odabasioglu1998prima}. Freezing the material
properties at a reference temperature $\theta$ gives the linearized
matrices $M=M(\theta)$ and $K=K(\theta)$. For frequency shift
$\sigma$, define
\begin{equation}
 A=(K+\sigma M)^{-1}M\,,\qquad
 B=(K+\sigma M)^{-1}G\,.
 \label{eq:mmblock}
\end{equation}
Repeated applications of $A$ to the input block $B$ generate the
block Krylov space:
\begin{equation}
 \mathcal K_{q}(A,B)=\operatorname{span}\{B,AB,\ldots,A^{q-1}B\}
 \label{eq:krylov}
\end{equation}
of order $q$. It locally matches moments at the reference temperature
$\theta$ and shift $\sigma$. 

Conventional moment matching uses one reference pair $(\theta,\sigma)$ and therefore represents dynamics with material properties frozen at a single temperature. Multipoint Krylov (MPK)
broadens this coverage by constructing blocks at $N_p$ reference
temperature--shift pairs and compressing their union using proper orthogonal decomposition (POD) to a
prescribed rank. It therefore incorporates material behavior at a
finite set of reference temperatures $\theta_i,\,i=1,\ldots,N_p$:
\begin{equation}
 V_{\mathrm{MPK}}=\operatorname{POD}\!\left[\mathcal K_{q_1}(A_1,B_1),\ldots,\mathcal K_{q_{N_p}}(A_{N_p},B_{N_p})\right]\,.
 \label{eq:mpk}
\end{equation}
However, this does not change the full-order
construction of the nonlinear matrices and vectors identified by
Eq.~\eqref{eq:rom}. Hyper-reduction is a solution to this bottleneck.

\subsection{Hyper-Reduction}
Hyper-reduction seeks to construct a nonlinear reduced matrix/vector from a
small number of sampled entries rather than from all $n$ DoFs. DEIM is one of the most popular hyper-reduction techniques ~\cite{chaturantabut2010deim}. For a nonlinear vector $f(V\widehat x)$, it first constructs
a basis $U_f\in\mathbb R^{n\times m}$ from nonlinear snapshots and then
selects $m$ interpolation entries through
$P_f\in\mathbb R^{n\times m}$ ($m\ll n$). The
full-order approximation and its reduced projection are:
\begin{align}
f(V\widehat x)
&\approx U_f(P_f^\top U_f)^{-1}P_f^\top f(V\widehat x)\,,\nonumber\\
V^\top f(V\widehat x)
&\approx C_fP_f^\top f(V\widehat x)\,,
\label{eq:deim}
\end{align}
where $C_f=V^\top U_f(P_f^\top U_f)^{-1}$ is pre-assembled in the offline stage. Thus only
the selected entries $P_f^\top f(V\widehat x)$ must be constructed
online. MDEIM extends the same interpolation principle to vectorized
matrices~\cite{wirtz2014deim}; in nonlinear thermal MOR, it can
approximate the heat-capacity and conductivity matrices, while DEIM
handles the leakage vector.

Sampling reduces the nonlinear construction cost, but it introduces both
offline and online overhead. DEIM and MDEIM require representative
nonlinear snapshots to construct $U_f$ and the corresponding matrix
bases. Moreover, constructing sampled finite-element entries still
requires recovering temperatures on the adjacent elements and
assembling their residual and Jacobian contributions. Although these methods alleviate the online computational bottleneck arising from nonlinearity, they rely on additional
full-order training data (which are expensive to obtain) and do not eliminate element-level online assembly. Our method \textit{SPLIT-MOR} does not need any training data and moves the element-level assembly burden fully to the offline stage. 

\section{\MakeUppercase{SPLIT-MOR}}
\label{sec:splitmor}

\begin{figure}[t]
  \centering
  \includegraphics[width=0.95\columnwidth]{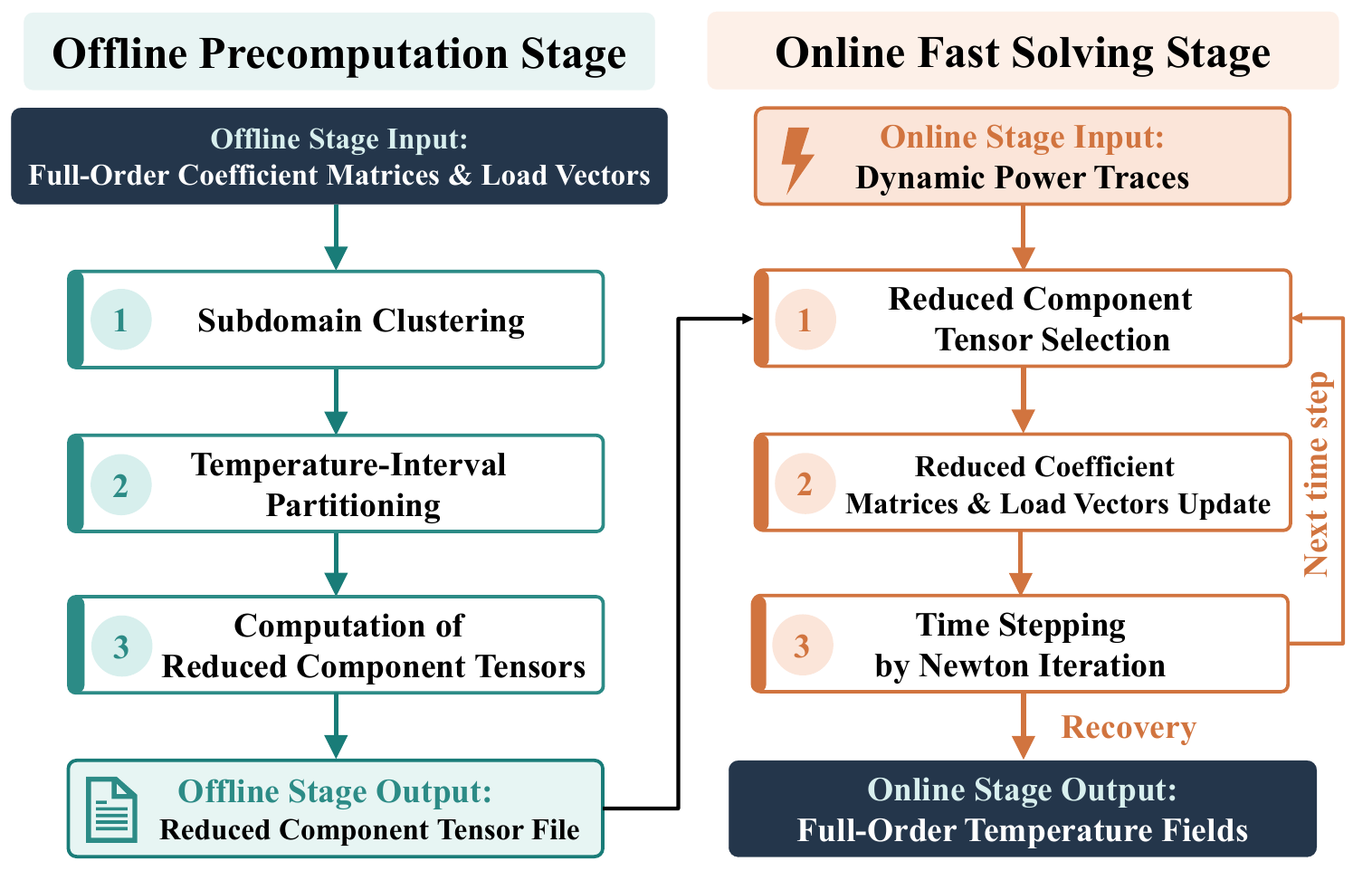}
  \caption{Offline pre-computation and online solving workflow of SPLIT-MOR. The offline
stage clusters subdomains, partitions the temperature range, and constructs
reduced component tensors; the online stage selects and updates the reduced coefficient matrices and load vectors using these tensors during Newton time stepping.}
  \label{fig:flow}
\end{figure}

\subsection{Framework Overview}
As introduced in Section \ref{sec: mor}, the remaining nonlinear MOR bottleneck is to evaluate
$\widehat M(\widehat x)$, $\widehat K(\widehat x)$, and
$\widehat l(\widehat x)$ without recovering the full state
$V\widehat x$. A single affine approximation over the complete
temperature range is inexpensive but inaccurate, whereas assigning an
independent approximation to every element preserves locality but
retains an online cost proportional to the mesh size. SPLIT-MOR
resolves this conflict by sharing one local affine model among elements
whose temperatures evolve similarly in the reduced space.

Figure~\ref{fig:flow} summarizes the resulting offline--online
decomposition. In the offline pre-computational stage, SPLIT-MOR groups elements using features
derived from $V$ into subdomains, partitions the temperature range under a fit
tolerance, and constructs reduced component tensors for each subdomain--interval pair.
In the online solving stage, it updates mean temperature
per subdomain, selects the corresponding interval, and combines pre-constructed reduced component tensors into global reduced coefficient matrices and load vectors. Thus all finite-element integration and
nonlinear-term evaluation are moved out of the online time-stepping loop.

The tensor viewpoint explains why a local affine representation is
useful. Let $a,b,c,d\in\{1,\ldots,n\}$ index full-order degrees of
freedom and let $\alpha,\beta,\gamma,\delta\in\{1,\ldots,r\}$ index
reduced coordinates. For example, a polynomial conductivity law gives
\begin{align}
K_{ab}(x)&=K^0_{ab}+K^1_{abc}x_c+K^2_{abcd}x_cx_d+\cdots,\nonumber\\
\widehat K_{\alpha\beta}(\widehat x)
&=\widehat K^0_{\alpha\beta}
+\widehat K^1_{\alpha\beta\gamma}\widehat x_\gamma
+\widehat K^2_{\alpha\beta\gamma\delta}
\widehat x_\gamma\widehat x_\delta+\cdots.
\label{eq:tensorsum}
\end{align}
The superscript on each coefficient tensor denotes its polynomial
order. Repeated indices are summed over their stated ranges. A degree-$q$
material model generates reduced tensors up to
order $q+2$, so storage grows as $O(r^{q+2})$. SPLIT-MOR does not assume
that the original law is polynomial. Instead, it keeps only the
zeroth- and first-order tensors of a piecewise-affine approximation;
local temperature intervals provide an accuracy guarantee while subdomain clustering keeps the number of online interval switches small.

\subsection{Offline Pre-Computation}
\begin{figure}[t]
  \centering
  \includegraphics[width=0.45\textwidth]{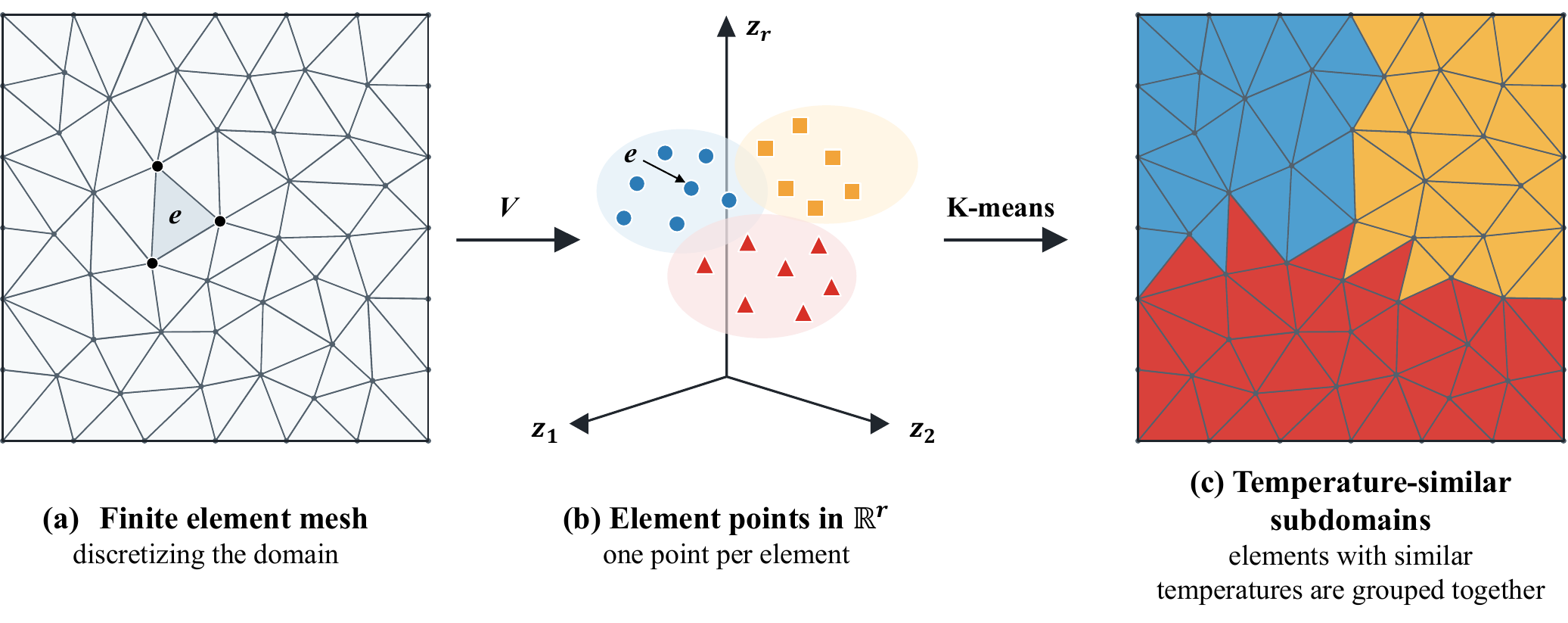}
  \caption{Reduced-basis feature clustering for constructing temperature-similar
subdomains. Each finite element is mapped to a point in $\mathbb R^r$, and
K-means groups nearby points into algebraic subdomains.}
  \label{fig:clustering}
\end{figure}

\paragraph {Subdomain clustering by reduced basis.}
Per-element interval labels would reintroduce $O(n)$ online work, so
SPLIT-MOR first groups elements with similar temperature
evolution as shown in Fig. \ref{fig:clustering}. Clustering precomputed temperature snapshots would operate
in a high-dimensional space whose dimension grows with the number of
time samples and would make SPLIT-MOR data-driven. Partitioning by
layout priors instead requires engineering knowledge and may reduce
generality. We therefore construct the clusters directly from the
reduced basis $V$.

Let $e$ index an element and let $\mathcal N_e$ be its vertex set.
The row average $v_e\in\mathbb R^r$ is the reduced feature of element
$e$, and $\theta_e$ is its characteristic element-average temperature
rise:
\begin{equation}
v_e=\frac{1}{|\mathcal N_e|}\sum_{a\in\mathcal N_e}V_{a,:}
\in\mathbb R^r,
\qquad \theta_e=v_e^\top\widehat x.
\label{eq:elementfeature}
\end{equation}
Let $N_s$ be the prescribed number of algebraic subdomains. K-means
partitions the features into $\{\Omega_s\}_{s=1}^{N_s}$, where $s$
is the subdomain index. A subdomain may contain
spatially disconnected elements; only additivity of the finite-element
integrals is required. For any two elements $e$ and $f$,
\begin{equation}
|\theta_e(t)-\theta_f(t)|
\leq\|v_e-v_f\|\,\|\widehat x(t)\|,
\label{eq:clusterbound}
\end{equation}
so feature proximity controls temperature mismatch. For SPLIT-MOR based on a non-data-driven MOR method, this method requires no snapshots. The
parameter $N_s$ controls the tradeoff between temperature coherence and
the number of stored component tensors.

\paragraph{Interval construction.}
Stronger material nonlinearity requires finer temperature intervals.
We therefore partition the admissible range of temperature rise
$[\theta_{\min},\theta_{\max}]$ by recursive bisection driven by the
affine-fit residual. Let $N_m$ be the number of materials. The index
$m=1,\ldots, N_m$ identifies a material, and $p$ identifies a candidate
temperature interval. Let $f_m$ denote one of the
heat-capacity, conductivity, or leakage laws of material $m$. On
interval $p$, SPLIT-MOR fits
$f_{m,p}(\theta)=f^0_{m,p}+f^1_{m,p}\theta$, where $f^0_{m,p}$ and
$f^1_{m,p}$ are its intercept and slope. The parameter $\epsilon$ is
the maximum permitted affine-fit residual after each property is
normalized by its range over $[\theta_{\min},\theta_{\max}]$.
If any residual exceeds $\epsilon$, the interval is bisected, and
both children are tested recursively. If $I$ intervals are accepted,
their $I+1$ boundaries are $\{\theta_p\}_{p=1}^{I+1}$. Their affine
coefficients are denoted by $c^j_{m,p}$, $k^j_{m,p}$, and
$p^j_{\mathrm{leak},m,p}$, where $j=0$ and $j=1$ identify the constant
and linear terms, respectively.

\paragraph{Reduced component-tensor assembly.}
Once the spatial and temperature partitions are fixed, all mesh
dependence can be precomputed. Let $\Omega_m$ be the region occupied by
material $m$. For $\mathbf r\in\Omega_m$, define the coefficient fields
$c^{jp}(\mathbf r)=c^j_{m,p}$,
$k^{jp}(\mathbf r)=k^j_{m,p}$, and
$p_{\rm leak}^{jp}(\mathbf r)=p^j_{\mathrm{leak},m,p}$. For subdomain
$s\in\{1,\ldots,N_s\}$ and interval $p\in\{1,\ldots,I\}$, assemble:
\begin{align}
\widehat M^{0sp}_{\alpha\beta}
&=\int_{\Omega_s}c^{0p}\phi_\alpha\phi_\beta \text{d}r, &
\widehat M^{1sp}_{\alpha\beta\gamma}
&=\int_{\Omega_s}c^{1p}\phi_\alpha\phi_\beta\phi_\gamma \text{d}r,\nonumber\\
\widehat K^{0sp}_{\alpha\beta}
&=\int_{\Omega_s}k^{0p}\nabla\phi_\alpha\!\cdot\!\nabla\phi_\beta \text{d}r, &
\widehat K^{1sp}_{\alpha\beta\gamma}
&=\int_{\Omega_s}k^{1p}\phi_\gamma
\nabla\phi_\alpha\!\cdot\!\nabla\phi_\beta\text{d}r ,\nonumber\\
\widehat l^{0sp}_{\alpha}
&=\int_{\Omega_s}p_{\rm leak}^{0p}\phi_\alpha\text{d}r, &
\widehat l^{1sp}_{\alpha\beta}
&=\int_{\Omega_s}p_{\rm leak}^{1p}\phi_\alpha\phi_\beta\text{d}r.
\label{eq:component}
\end{align}
Here $\widehat M^{0sp}$ and $\widehat K^{0sp}$ are $r\times r$
matrices; $\widehat M^{1sp}$ and $\widehat K^{1sp}$ are
$r\times r\times r$ tensors; and $\widehat l^{0sp}$ and
$\widehat l^{1sp}$ are a length-$r$ vector and an $r\times r$ matrix,
respectively. These component vectors/matrices/tensors map
the affine material laws directly to reduced coefficient matrices and load vectors and remove all
quadrature and local-to-global assembly from the online stage.

Finally, SPLIT-MOR preassembles:
\begin{equation}
\Gamma_s^0=\int_{\Omega_s}1\text{d}r,\qquad
\Gamma^1_{s\alpha}=\int_{\Omega_s}\phi_\alpha\text{d}r,\qquad
\theta_s=\frac{\Gamma^1_{s\alpha}\widehat x_\alpha}{\Gamma_s^0}.
\label{eq:average}
\end{equation}
Here $\theta_s$ is the volume-average temperature rise of subdomain
$s$. Because $\Gamma_s^0$ and $\Gamma^1_{s\alpha}$ are preassembled,
the online phase evaluates $\theta_s$ and selects its temperature
interval without element-wise integration.

The cost is $O(nIr^3)$ if the integrations are repeated for all $I$
intervals. When the property laws are specified for $N_m$ discrete
materials and $N_m<I$, a more efficient strategy is to assemble
interval-independent geometry tensors once. Let
$\Omega_{s,m}=\Omega_s\cap\Omega_m$ be the part of subdomain $s$
occupied by material $m$. The second- and third-order conductivity
geometry tensors are:
\begin{equation}
\mathcal K^{2sm}_{\alpha\beta}
=\int_{\Omega_{s,m}}\nabla\phi_\alpha\cdot\nabla\phi_\beta\text{d}r,
\quad
\mathcal K^{3sm}_{\alpha\beta\gamma}
=\int_{\Omega_{s,m}}\phi_\gamma
\nabla\phi_\alpha\cdot\nabla\phi_\beta\text{d}r.
\label{eq:geometry}
\end{equation}
The interval-dependent component matrices/tensors follow from weighted sums such as
$\widehat K^{jsp}=\sum_m k^j_{m,p}\mathcal K^{(j+2)sm}$; analogous
geometry tensors produce $\widehat M^{jsp}$ and $\widehat l^{jsp}$. Reusing these geometry tensors across
intervals reduces offline assembly from $O(nIr^3)$ to a single $O(nr^3)$ pass,
with $O(N_sIr^3)$ storage for interval components to $O(N_sN_mr^3)$ for
material-factored tensors.

\begin{figure}[h]
  \centering
  \includegraphics[width=0.95\columnwidth]{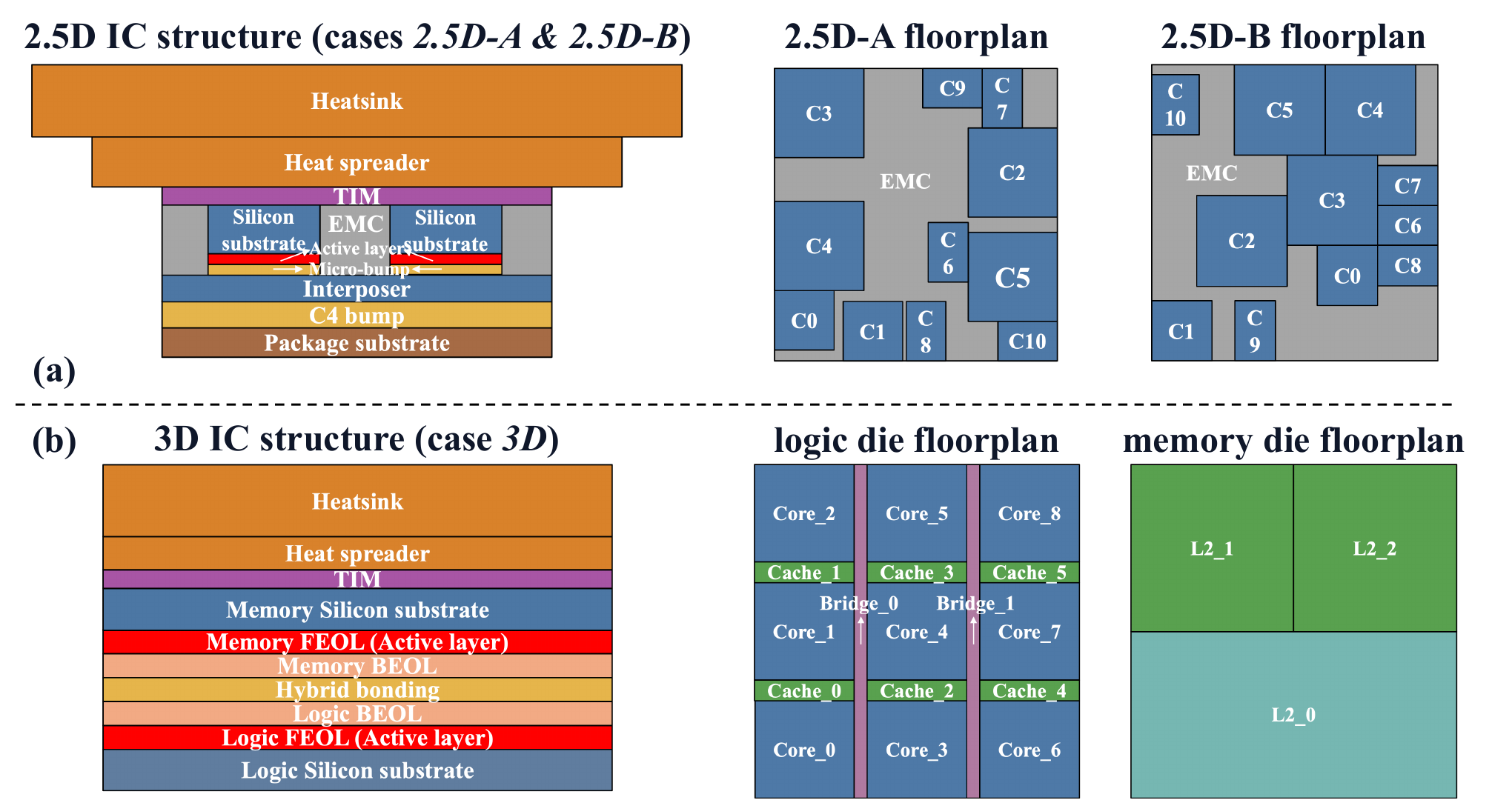}
  \caption{Layer stacks and active-layer floorplans of the three test cases.
The 2.5D-A and 2.5D-B cases share the same stack but use different chiplet
placements; the 3D case uses face-to-face logic and memory dies.}
  \Description{The top row shows the 2.5D package stack and two chiplet floorplans; the bottom row shows the 3D stack and its logic and memory floorplans.}
  \label{fig:packages}
\end{figure}

\providecommand{\NA}{\textsc{N/A}}

\begin{table*}[ht]
\centering
\caption{Summarized accuracy and runtime for the three test cases. Parenthesized values
report speedup relative to MDEIM for total offline time and relative to
COMSOL for online time.}
\label{tab:results}
\small
\renewcommand{\arraystretch}{0.92}
\setlength{\tabcolsep}{4.5pt}
\begin{tabular*}{0.98\textwidth}{@{\extracolsep{\fill}}cccccc@{}}
\toprule
Cases & Methods & Total offline & Online & Avg. RMSPE & MAE\\
\midrule
\multirow{5}{*}{2.5D-A}
& COMSOL (Ground truth) & \NA & \qty{15}{\minute} \qty{51.29}{\second} (\num{1.00}$\times$) & \qty{0}{\percent} & \qty{0}{\degreeCelsius}\\
& LINEAR-COMSOL & \NA & \qty{8}{\minute} \qty{37.06}{\second} (\num{1.84}$\times$) & \qty{5.9241}{\percent} & \qty{15.9070}{\degreeCelsius}\\
& MDEIM & \qty{27}{\minute} \qty{55.80}{\second} (\num{1.00}$\times$) & \qty{1}{\minute} \qty{2.79}{\second} (\num{15.15}$\times$) & \qty{0.0800}{\percent} & \qty{0.5344}{\degreeCelsius}\\
& SPLIT-MOR w/o MPK & \qty{1}{\minute} \qty{45.00}{\second} (\num{15.96}$\times$) & \qty{18.75}{\second} (\num{50.74}$\times$) & \qty{0.1692}{\percent} & \qty{1.7427}{\degreeCelsius}\\
& \textbf{SPLIT-MOR (Ours)} & \textbf{2 min 39.00 s} (\textbf{10.54$\times$}) & \textbf{19.87 s} (\textbf{47.88$\times$}) & \textbf{0.1033 \%} & \textbf{0.6576 $^\circ$C}\\
\addlinespace[2pt]

\multirow{5}{*}{2.5D-B}
& COMSOL (Ground truth) & \NA & \qty{91}{\minute} \qty{50.47}{\second} (\num{1.00}$\times$) & \qty{0}{\percent} & \qty{0}{\degreeCelsius}\\
& LINEAR-COMSOL & \NA & \qty{44}{\minute} \qty{24.58}{\second} (\num{2.07}$\times$) & \qty{6.4342}{\percent} & \qty{19.4665}{\degreeCelsius}\\
& MDEIM & \qty{104}{\minute} \qty{3.00}{\second} (\num{1.00}$\times$) & \qty{1}{\minute} \qty{20.04}{\second} (\num{68.85}$\times$) & \qty{0.0677}{\percent} & \qty{0.6250}{\degreeCelsius}\\
& SPLIT-MOR w/o MPK & \qty{6}{\minute} \qty{33.60}{\second} (\num{15.86}$\times$) & \qty{19.76}{\second} (\num{278.87}$\times$) & \qty{0.1613}{\percent} & \qty{2.3614}{\degreeCelsius}\\
& \textbf{SPLIT-MOR (Ours)} & \textbf{9 min 58.80 s} (\textbf{10.43$\times$}) & \textbf{21.00 s} (\textbf{262.40$\times$}) & \textbf{0.0997 \%} & \textbf{0.8536 $^\circ$C}\\
\addlinespace[2pt]

\multirow{5}{*}{3D}
& COMSOL (Ground truth) & \NA & \qty{87}{\minute} \qty{38.98}{\second} (\num{1.00}$\times$) & \qty{0}{\percent} & \qty{0}{\degreeCelsius}\\
& LINEAR-COMSOL & \NA & \qty{37}{\minute} \qty{2.90}{\second} (\num{2.37}$\times$) & \qty{19.1160}{\percent} & \qty{28.1944}{\degreeCelsius}\\
& MDEIM & \qty{300}{\minute} \qty{11.30}{\second} (\num{1.00}$\times$) & \qty{47.66}{\second} (\num{110.34}$\times$) & \qty{0.0242}{\percent} & \qty{0.2549}{\degreeCelsius}\\
& SPLIT-MOR w/o MPK & \qty{2}{\minute} \qty{52.92}{\second} (\num{104.16}$\times$) & \qty{6.36}{\second} (\num{826.88}$\times$) & \qty{0.0487}{\percent} & \qty{0.3213}{\degreeCelsius}\\
& \textbf{SPLIT-MOR (Ours)} & \textbf{11 min 3.60 s} (\textbf{27.14$\times$}) & \textbf{6.34 s} (\textbf{829.49$\times$}) & \textbf{0.0270 \%} & \textbf{0.1746 $^\circ$C}\\
\bottomrule
\end{tabular*}
\end{table*}

\begin{figure*}[!t]
  \centering
  \includegraphics[
  width=0.75\textwidth,
  ]{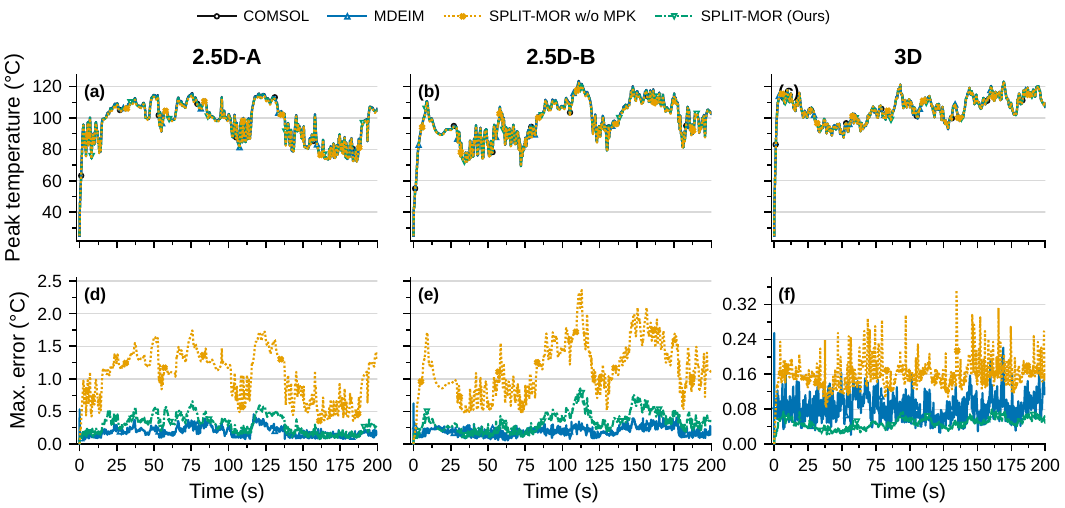}
  \caption{Peak-temperature and maximum absolute-error histories over the 200-s
unseen-load tests. Columns correspond to 2.5D-A, 2.5D-B, and 3D; the top row
compares peak temperatures, and the bottom row reports maximum nodal errors
relative to COMSOL.}
  \Description{Six plots compare COMSOL, MDEIM, SPLIT-MOR (ours), and SPLIT-MOR w/o MPK. The top row shows peak-temperature histories, and the bottom row shows maximum absolute-error histories for the three packages.}
  \label{fig:curves}
\end{figure*}

\begin{figure}[!t]
  \centering
  \includegraphics[width=0.75\columnwidth]{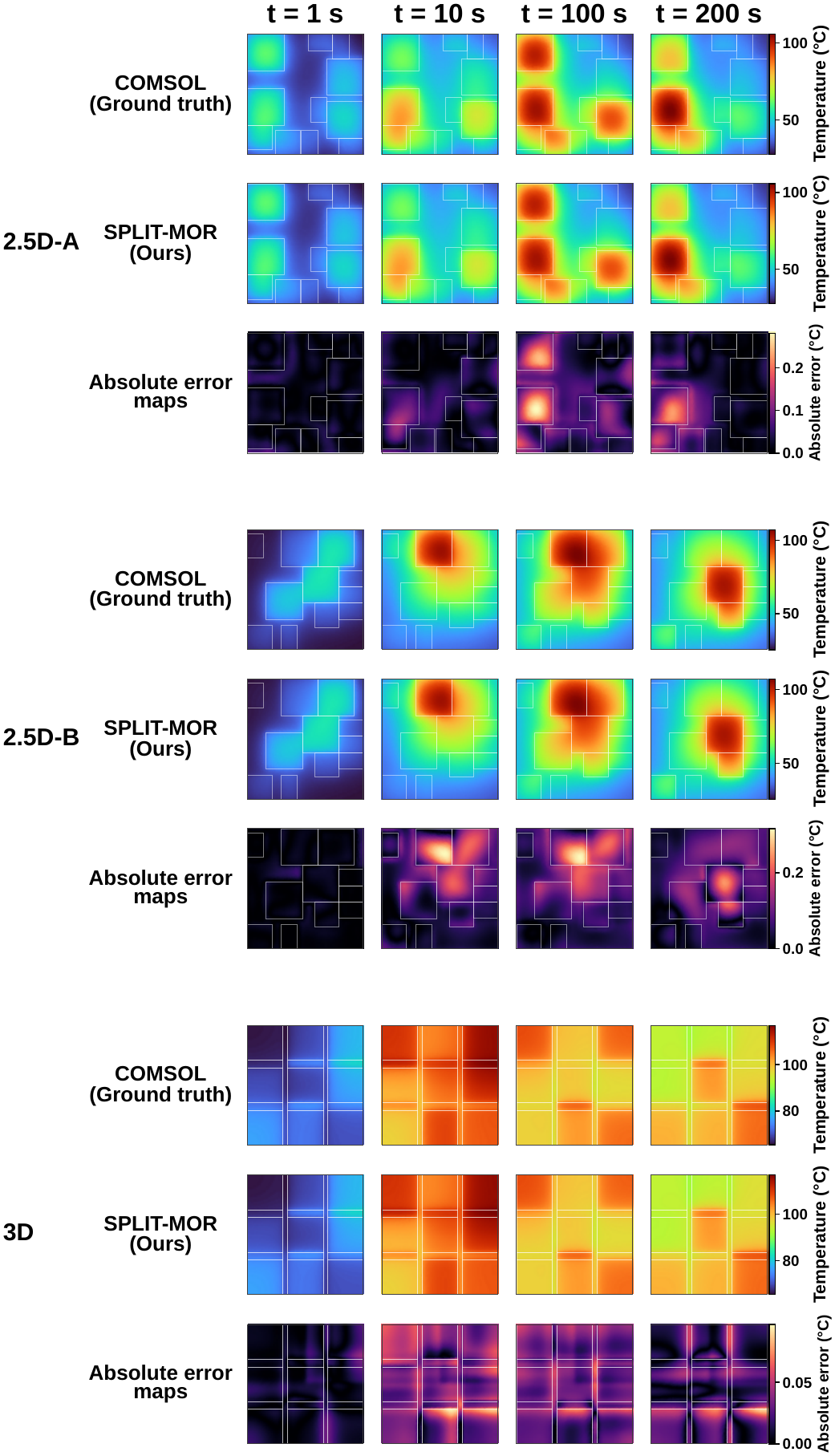}
  \caption{Temperature and absolute-error maps at \qtylist{1;10;100;200}{\second} on the active layers of the 2.5D cases and on the logic die of the 3D case.}
  \Description{Six plots compare COMSOL, MDEIM, SPLIT-MOR (ours), and SPLIT-MOR w/o MPK. The top row shows Peak-temperature histories and the bottom row shows maximum absolute-error histories for the three packages.}
  \label{fig:active_layer}
\end{figure}

\subsection{Online Reduced Solver}
At each Newton iteration, $p_s\in\{1,\ldots,I\}$ denotes the active
interval label of subdomain $s$ and is selected such that
$\theta_{p_s}\leq\theta_s<\theta_{p_s+1}$. Additivity gives:
\begin{equation}
\widehat K^j=\sum_s\widehat K^{jsp_s},\quad
\widehat M^j=\sum_s\widehat M^{jsp_s},\quad
\widehat l^j=\sum_s\widehat l^{jsp_s},\quad j\in\{0,1\}.
\label{eq:globalsum}
\end{equation}
These global component vectors/matrices/tensors and the temperature-interval number of each subdomain are cached because, in most time steps, only a few subdomains undergo a temperature-interval transition. This cached-update strategy can, therefore, significantly accelerate online computation. Let $A$ stand for one of $M$, $K$, or $l$, and let $\widehat A^{jsp}$ denote its stored
component for the affine order $j$, subdomain $s$ and interval $p$. If
the subdomain $s$ changes from $p_s^{\rm old}$ to $p_s^{\rm new}$, the
corresponding cache is updated: 
\begin{equation}
\widehat A^j\leftarrow\widehat A^j
-\widehat A^{js p_s^{\rm old}}
+\widehat A^{js p_s^{\rm new}},
\qquad A\in\{M,K,l\},
\label{eq:cacheupdate}
\end{equation}
instead of recomputing Eq.~\eqref{eq:globalsum}. The resulting reduced nonlinear system is solved by Newton iteration
with backtracking. Overall, evaluating all
subdomain averages costs $O(N_sr)$; updating $N_{\rm ch}$ switched subdomains
and solving the reduced system cost $O(N_{\rm ch}r^3)$ and $O(r^3)$,
respectively, so the online complexity is independent of $n$.

\section{\MakeUppercase{Experiments}}
\label{sec:experiments}

\subsection{Setup}
This section studies two 2.5D chiplet packages and one 3D chiplet package. The two 2.5D packages share the same structure except for the placement of their active chiplets. Together, the three packages allow us to evaluate the applicability, efficiency, and accuracy of SPLIT-MOR across different 2.5D/3D IC structures. Their layer stacks and active-layer floorplans are shown in Fig.~\ref{fig:packages}.

The compared methods are: (i) COMSOL~\cite{comsol}, commercial software serving as the ground truth; (ii) LINEAR-COMSOL, which freezes all material properties and leakage power at room temperature and neglects nonlinearity, as an ablation demonstrating the need to retain nonlinear terms; (iii) MDEIM, the state-of-the-art hyper-reduction method for offline and online efficiency comparison; (iv) SPLIT-MOR w/o MPK, which applies SPLIT-MOR with a moment-matching basis constructed only at room temperature, as an ablation testing whether MPK reduces error by accounting for material properties at multiple temperatures; and (v) SPLIT-MOR, our method. There are many academic thermal simulators such as HotSpot~\cite{zhang2015hotspot}, Pact~\cite{yuan2022pact}, MTA~\cite{ladenheim2018mta}, 3D-ICE~\cite{11539294}, and ATSim~\cite{wang2025atsim3}. However, they are all full-order simulators like COMSOL, and none of them support nonlinear leakage power, thermal conductivity, and heat capacity at the same time, so we did not compare with these academic thermal simulators.

For a fair comparison, all methods for a given package use the same Gmsh mesh~\cite{geuzaine2009gmsh}. The COMSOL model and non-COMSOL model of the same structure share similar mesh densities and total number of DoFs. DOLFINx~\cite{BarattaEtal2023} is used for DoF management, while NumPy/SciPy is used for vectorized P1 assembly in the offline stages and dense algebra in the online stage of SPLIT-MOR and MDEIM. All experiments run on an Intel Xeon Gold 6248R Linux server. COMSOL, offline stages, I/O, and full-field recovery use \num{24} cores, whereas the online computation uses one core. Ambient and initial temperatures are $\qty{25}{\degreeCelsius}$, and all transient solvers use backward Euler with $\Delta t=\qty{0.05}{\second}$.

MPK samples temperature rises $\{\num{0}, \num{25}, \num{50}, \num{75}, \num{100}, \num{150}\}$ $^\circ$C and frequency shifts $\{0,0.005,0.05,0.5,2,20\}\unit{\per\second}$. For all cases, the training set of MDEIM consists of 36 independent \qty{10}{\second} full-order trajectories simulated by COMSOL. Within each trajectory, the dynamic power at every power port is sampled independently and randomly every \qty{0.5}{\second} and linearly interpolated between consecutive samples. The corresponding temperature results are used to construct the snapshots required for MDEIM training. The test dynamic power traces for all of the methods are the same and last for \qty{200}{\second} (4000 steps). During testing, every power port normally follows the same random interpolation process, but at every 0.5 s update it also has an independent \qty{1}{\percent} probability of entering either a 0.9 high-load mode or a 0.1 low-load mode for \qtyrange{2.5}{30}{\second}. This setting increases the diversity of the test dynamic power traces.

As shown in Fig. \ref{fig:packages}(a), the 2.5D IC comprises a heat sink, heat spreader, thermal interface material (TIM), active chiplets, silicon interposer/C4, and substrate. The floorplans of the chiplets are from the placement results of \cite{wang2024atplace2}. On the heat-sink top, $h=\qty{1e4}{\watt\per\square\metre\per\kelvin}$; on the substrate bottom, $h=\qty{100}{\watt\per\square\metre\per\kelvin}$. All other surfaces are adiabatic. The heat capacities and thermal conductivities of all materials and the leakage powers of all active layers are temperature dependent. The 2.5D-A mesh has $5.5\times10^4$ DoFs and $2.7\times10^5$ tetrahedra, whereas the 2.5D-B mesh has $1.98\times10^5$ DoFs and $1.08\times10^6$ tetrahedra. For both cases, the MDEIM sample counts are 40, the SPLIT-MOR ranks are 80, SPLIT-MOR uses 32 subdomains, and the interval-fit tolerance is \qty{0.1}{\percent}.

As shown in Fig. \ref{fig:packages}(b), the 3D IC comprises a logic die and a memory die connected face-to-face with hybrid bonding~\cite{wang2026etla}. The heat-sink top uses $h=\qty{8000}{\watt\per\square\metre\per\kelvin}$, and all other surfaces are adiabatic. The heat capacity and conductivity of every material depend on temperature. Leakage in both active layers is also temperature dependent. Its mesh has \num{2.23e5} DoFs and \num{1.26e6} tetrahedra. The MDEIM sample count is 24, the SPLIT-MOR rank is 50, SPLIT-MOR uses 24 subdomains, and the interval-fit tolerance is \qty{0.04}{\percent}.

\subsection{Results}
We first provide detailed definitions of all accuracy and runtime metrics reported in the Results section. Average error is the time average of the spatial root-mean-square percentage error (RMSPE), and maximum absolute error (MAE) is the largest absolute nodal error over all time steps; both use COMSOL as the reference. The total offline time is the wall-clock time required for the entire offline procedure, including training-set preparation and MOR. Online time excludes I/O and full-field recovery so that it reflects the cost of advancing the reduced system. Table~\ref{tab:results} reports the summarized metrics, while Fig.~\ref{fig:curves} shows how the peak temperature and nodal error evolve throughout the 200 s tests. Fig.~\ref{fig:active_layer} presents the temperature and absolute-error maps at \qtylist{1;10;100;200}{\second}, extracted from the active layers of the two 2.5D cases and from the logic die of the 3D case.

Comparing COMSOL with LINEAR-COMSOL in Table~\ref{tab:results} shows that material nonlinearity and leakage power cannot be neglected. Freezing the properties at room temperature produces $\qtyrange{5.9}{19.1}{\percent}$ average error and underestimates the peak temperature by $\qtyrange{15.9}{28.2}{\degreeCelsius}$. Such a temperature deficit would seriously mislead designers and automatic DTM strategies.

Compared with the COMSOL reference, our SPLIT-MOR achieves online speedups of $\qtyrange{47.88}{829.49}{\times}$ while maintaining average errors of only $\qtyrange{0.0270}{0.1033}{\percent}$ and maximum errors of 0.1746 $^\circ$C to 0.8536 $^\circ$C. Figure~\ref{fig:curves} further shows that its predicted peak-temperature trajectories closely track the COMSOL solutions over the extended 200 s test horizon. SPLIT-MOR attains accuracy comparable to MDEIM without requiring the costly full-order training datasets used by data-driven reduction methods, thereby achieving total-offline speedups of $10.43$--$27.14\times$ over MDEIM.

SPLIT-MOR remains $3.16$--$7.52\times$ faster than MDEIM despite using an MPK rank approximately twice the MOR rank, because constructing the sampled finite-element entries for MDEIM still requires recovering temperatures on adjacent elements and assembling their residual and Jacobian contributions online (\qty{87.2}{\percent} of its online computational time for case 2.5D-A). In contrast, SPLIT-MOR avoids such element-level assembly and further reduces its online cost through dynamic caching. To quantify this contribution, we rerun SPLIT-MOR with caching disabled, forcing the global reduced tensors to be recomputed using all subdomains at every Newton iteration. While producing identical temperature results, the cache-disabled version takes $\num{106.71} \times$, $\num{122.54}\times$, and $\num{41.47}\times$ as much online time as the current cached version for 2.5D-A, 2.5D-B, and 3D, respectively.

The comparison with SPLIT-MOR w/o MPK isolates the contribution of the MPK basis. At the same reduced rank, SPLIT-MOR decreases the average error by $\qty{38.9}{\percent}$, $\qty{38.2}{\percent}$, and $\qty{44.6}{\percent}$ for 2.5D-A, 2.5D-B, and 3D cases, respectively. The corresponding reductions in maximum error are $\qty{62.3}{\percent}$, $\qty{63.9}{\percent}$, and $\qty{45.7}{\percent}$. SPLIT-MOR w/o MPK uses a moment-matching basis constructed at a single room-temperature material state, whereas the MPK basis incorporates reference pairs at multiple temperature shifts. The ablation results show that this multi-temperature construction improves the representation of temperature-dependent dynamics, albeit with a larger basis-construction cost.

\FloatBarrier

\section{\MakeUppercase{Conclusion}}
\label{sec:conclu}
In this work, we propose SPLIT-MOR, a novel nonlinear MOR method that removes full-order online work from nonlinear transient thermal MOR by clustering reduced-basis features, fitting material laws over temperature intervals, and assembling subdomain affine tensors offline. It imposes no specific nonlinear property law and does not rely on expensive full-order training data like previous methods. Experiments on three 2.5D/3D packages demonstrate that its online time reaches up to an $830\times$ speedup over COMSOL, a $7.5\times$ speedup over MDEIM, with a $27\times$ offline time speedup over MDEIM, while keeping average error below $0.11\%$ and maximum error below 0.86 $^\circ$C. Future work will focus on further improving the offline and online efficiency of our method.

\begin{acks}
The work was supported in part by Beijing High Innovation Plan (202604841450) and Jiangsu National Science and Technology Major Project (BG2025014).
\end{acks}

\clearpage
{
\bibliographystyle{ACM-Reference-Format}
\bibliography{./ref/MOR,./ref/common_thermal,./ref/Software}
}

\end{document}